\documentstyle[12pt]{article}

\begin{document}

\begin{titlepage}

\title{\large\bf CP Violation, Sneutrino Oscillation 
and Neutrino Masses in R-parity Violating Supersymmetric
Standard Model}

\author{Eung Jin Chun \\
 {\normalsize\it  Department of Physics and Center for Theoretical Physics}, \\
 {\normalsize\it  Seoul National University, Seoul 151-747, Korea}  
}


\date{}
\maketitle

\begin{abstract} {\normalsize
In supersymmetric theories, sneutrino--anti-sneutrino mixing  
can occur with the oscillation time $\sim 0.01$ ps corresponding
the atmospheric neutrino mass scale $\sim 0.05$ eV.
We explore the possibility of observing sneutrino oscillation 
phenomena and CP violation when R-parity violation explains the
observed neutrino masses and mixing. It is shown for some
parameter region in the bilinear model of R-parity violation
that the tiny sneutrino mass splitting and time-dependent 
CP violating asymmetries  could be measured in the future experiments  
if the tau sneutrino is the lightest supersymmetric particle.
} \end{abstract}



\thispagestyle{empty}
\end{titlepage}


The atmospheric neutrino data from the Super-Kamiokande experiment 
\cite{skatm} strongly suggest the presence of sub-eV neutrino masses.
As is commonly  accepted,  such tiny masses are most likely to be 
of the Majorana type breaking lepton number by two units ($\Delta L=2$). 
In supersymmetric theories where each neutrino species $\nu$ is
accompanied by a complex scalar $\tilde{\nu}$, similar lepton number 
violation should occur in the scalar (sneutrino)  sector.   
The sneutrinos $\tilde{\nu}$ and $\tilde{\nu}^*$
generally have the mass terms,
\begin{equation} \label{snumaj}
 -{\cal L}=  m_{\tilde{\nu}}^2 \tilde{\nu}\tilde{\nu}^* 
 +{1\over2} \left(m^2_M \tilde{\nu}\tilde{\nu} + h.c.  \right)  \,,
\end{equation}
where $m^2_{\tilde{\nu}}$ is the usual slepton mass of the order
100 GeV and $m^2_M$ carries the same lepton number 
as the Majorana neutrino mass $m_\nu$. 
As we will see, the soft $B$-terms are the origin of 
the mass $m_M^2$ in our case.
This `Majorana' sneutrino mass term generates  mass splitting 
and mixing in the sneutrino-anti-sneutrino system \cite{hirsch,grossman}.
One can easily find from Eq.~(1) that the sneutrino mass-squared
eigenvalues are $m_{\tilde{\nu}}^2 \pm |m^2_M|$.
In general supersymmetric models, one finds 
$m_M^2 \sim m_\nu m_{\tilde{\nu}}$ and thus
the sneutrino mass splitting of the order of the neutrino mass;
$\Delta m_{\tilde{\nu}} \simeq |m^2_M|/m_{\tilde{\nu}} \sim m_\nu$.
Similarly to the case of neutrinos,
such a small mass difference could only  be measured through the observation 
of sneutrino oscillation, providing a novel opportunity to probe 
$\Delta L=2$ lepton number violating phenomena 
in the future collider experiments \cite{grossman}.

Mixing phenomena in the sneutrino--anti-sneutrino system are analogous
to those in the $B$--$\bar{B}$ system for which  we refer the readers 
to  the reviews in  Ref.~\cite{pdg}.  
Extending the original investigation in Ref.~\cite{grossman},
we wish to formulate CP violating effects in 
time-dependent sneutrino oscillation  and apply them to a specific
model of neutrino masses, namely, supersymmetric standard model with
R-parity violation.  For this, we assume the neutrino masses
explaining the Atmospheric neutrino data with
the largest neutrino mass $m_{\nu_3}\sim 0.05$ eV  and 
the large $\nu_\mu$--$\nu_\tau$ mixing \cite{skatm}.  
It is amusing to notice that the neutrino mass $m_{\nu_3}$ corresponds 
to 
the proper sneutrino oscillation time,
$ t_{osc}= \Delta m_{\tilde{\nu}}^{-1} =
{ m_{\tilde{\nu}} \over p} L \sim 0.013$ ps where $L$ measures the 
distance between the sneutrino production and decay vertex 
and $ m_{\tilde{\nu}} \over p$ is the boost factor of a few.  
Therefore, considering  a spatial resolution of a few hundredths 
of ps \cite{btev}, the atmospheric neutrino mass scale could be directly 
probed by observing time-dependent sneutrino oscillation in the
future experiments.

For sneutrino oscillation to occur, it is required that 
the sneutrino life-time is longer than its oscillation time; 
$$ \Gamma_{\tilde{\nu}} <  \Delta  m_{\tilde{\nu}} \,.$$ 
In the supersymmetric see-saw model considered in Ref.~\cite{grossman},
the sneutrino mass splitting is given by $\Delta  m_{\tilde{\nu}}/m_\nu
\approx 2(A+\mu\cot\beta)/m_{\tilde{\nu}}$ where $A$ is the  trilinear
soft mass and $\tan\beta$ is the ratio of the vacuum expectation values
of two Higgs bosons.  With $m_\nu \sim 0.05$ eV, one generically expects 
$\Delta  m_{\tilde{\nu}} < 1$ eV.  Then, the requirement
$\Gamma_{\tilde{\nu}} <  1$ eV can be arranged in a limited region of
parameter space under the assumption that the sneutrino has only 
three-body decay channels and the stau $\tilde{\tau}_R$ is 
the ordinary lightest supersymmetric particle (LSP) \cite{grossman}. 
In this scheme, 
CP violating effects  could arise if a nontrivial phase exists in the 
`Majorana' sneutrino mass.

Another interesting scheme for generating neutrino masses  and mixing
is to allow R-parity violation in the supersymmetric 
standard model \cite{hall}.  
In this case, the LSP is unstable and decays through 
R-parity violating couplings which generate both 
neutrino masses and sneutrino mass splittings.  Therefore, 
it is expected that R-parity violation can naturally lead to  
the suppressed decay rate and observable sneutrino oscillation 
phenomena {\it if a sneutrino is the LSP}.
Whether or not a sneutrino can be the LSP depends on the models of
supersymmetry breaking.  In the constrained framework of minimal
supergravity \cite{wells} or gauge-mediated supersymmetry
breaking \cite{gmsb}, the sneutrino LSP may be obtained 
in a limited parameter space where the D-terms can 
make $m^2_{\tilde{\nu}}$ smaller than $m^2_{\tilde{\tau}_R}$.
But, such a possibility appears to be ruled out \cite{hebb} by current
experiments on the invisible $Z$ width, providing the limit
$ m_{\tilde{\nu}} > 44.7$ GeV \cite{pdg}.   Therefore,
we assume less constrained (non-universal) soft masses so that
slepton doublets are  lighter  than slepton singlets and thus
sneutrinos lighter than charged sleptons.   Most favorable framework 
for this is SU(5) grand unification theory
where the common soft mass of ${\bf \bar{5}}$ multiplets 
is smaller than that of ${\bf 10}$ multiplets or gauginos.
Let us remark that our discussion can be complimentary to the
case of a neutralino LSP in testing the R-parity violation 
model of neutrino masses and mixing \cite{vissani}.

\medskip

We begin our main discussion by considering the general 
sneutrino-fermion-fermion couplings as follows:
\begin{equation} \label{snuff}
 {\cal L} = -{g \over \sqrt{2}} \; \tilde{\nu}_i\;  \overline{f} 
 \left[ P_L K_{L,i}^{f \bar{f}'} + P_R K_{R,i}^{f \bar{f}'} \right] f'  + h.c.
\end{equation}
where $P_{L,R}=(1\mp \gamma_5)/2$ and the index $i$ refers to the lepton 
flavor.  Neglecting mixing between different sneutrino flavors,
each sneutrino species $\tilde{\nu}_i$ has the mass term (\ref{snumaj})
with complex $m^2_M$.
The mass eigenstates denoted by two real fields, $\tilde{\nu}_{i1}$ and 
$\tilde{\nu}_{i2}$,  have the following  decay widths:
\begin{eqnarray} \label{dwidths}
\Gamma_{\tilde{\nu}_i} \equiv \Gamma_{\tilde{\nu}_{i1}}
 +\Gamma_{\tilde{\nu}_{i2}} = {g^2\over 16\pi} \sum_{f,f'} 
\left[ |K_{L,i}^{f\bar{f}'}|^2 + |K_{R,i}^{f\bar{f}'}|^2 \right] \\
\Delta\Gamma_{\tilde{\nu}_i} \equiv \Gamma_{\tilde{\nu}_{i1}}
  -\Gamma_{\tilde{\nu}_{i2}} = {g^2\over 16\pi} \sum_{f,f'} 
\left[ K_{L,i}^{f\bar{f}'}K_{R,i}^{f'\bar{f}}e^{i\phi_M} + c.c. \right]
\end{eqnarray}
where $\phi_M\equiv {\rm Arg}(m^2_M)$.
The time-evolution of the state identified as the sneutrino $\tilde{\nu}_i$
or the anti-sneutrino $\tilde{\nu}^*_i$
at an initial time $t=0$ is given respectively by
\begin{eqnarray} \label{snuosc}
 |\tilde{\nu}_i(t)\rangle  &=&
         g_+(t) |\tilde{\nu}_i\rangle
         + g_-(t) e^{-i\phi_M} |\tilde{\nu}^*_i\rangle \nonumber \\
 |\tilde{\nu}^*_i(t)\rangle &=&
          g_+(t) |\tilde{\nu}^*_i\rangle 
         + g_-(t) e^{+i\phi_M} |\tilde{\nu}_i\rangle
\end{eqnarray}
where $g_\pm(t)={1\over2}\exp(-{1\over2}\Gamma_{\tilde{\nu}_{i1}}t
-i m_{\tilde{\nu}_{i1}}t) [1\pm\exp({1\over2}\Delta\Gamma_{\tilde{\nu}_i}t
+ i\Delta m_{\tilde{\nu}_i} t)]$.
Then, the time-dependent CP asymmetry 
for the sneutrino decay to the final state $f\bar{f'}$ is
\begin{eqnarray} \label{Acpt}
A^i_{CP}(f\bar{f'}; t) &\equiv& 
      { \Gamma(\tilde{\nu}_i(t) \to f \bar{f'}) 
                    -  \Gamma(\tilde{\nu}^*_i(t) \to f \bar{f'})
       \over  \Gamma(\tilde{\nu}_i(t) \to f \bar{f'}) +
              \Gamma(\tilde{\nu}^*_i(t) \to f \bar{f'}) } \\
 &=& 
 { \cos(\Delta m_{\tilde{\nu}_i}t)  X_i^{ff'} 
    -2 \sin(\Delta m_{\tilde{\nu}_i}t)\, {\rm Im}(Y_i^{ff'}) \over
 c(t) - s(t) \,{\rm Re}(Y_i^{ff'}) }
 \nonumber
\end{eqnarray}
where  $c(t) =  (e^{\Delta\Gamma_{\tilde{\nu}_i}t/2}
      +e^{-\Delta\Gamma_{\tilde{\nu}_i}t/2})/2$,
$s(t) = (e^{\Delta\Gamma_{\tilde{\nu}_i}t/2}
      -e^{-\Delta\Gamma_{\tilde{\nu}_i}t/2})$ and 
\begin{eqnarray}
 X_i^{ff'} &\equiv &
   { |K_{L,i}^{f\bar{f}'}|^2 +  |K_{R,i}^{f'\bar{f}}|^2 
      - (f \leftrightarrow f') \over 
 |K_{L,i}^{f\bar{f}'}|^2 + |K_{R,i}^{f'\bar{f}}|^2 
       + (f \leftrightarrow f') } \,, \nonumber\\
 Y_i^{ff'} &\equiv & 
     {   e^{i\phi_M}  K_{L,i}^{f\bar{f}'}K_{R,i}^{f'\bar{f}} +
               (f \leftrightarrow f') \over 
 |K_{L,i}^{f\bar{f}'}|^2 + |K_{R,i}^{f'\bar{f}}|^2 + (f \leftrightarrow f') } .
\end{eqnarray}
For the `flavor-specific' final state $f\bar{f}'$ arising from 
the decay of $\tilde{\nu}_i$, that is, 
$\tilde{\nu}_i \rightarrow f \bar{f'} 
\leftarrow\hspace{-2ex}/\hspace{1ex}  \tilde{\nu}_i^*$,
we have $X_i^{ff'}=1$ and $Y_i^{ff'}=0$.  
Thus, the CP asymmetry in Eq.~(\ref{Acpt}) becomes 
\begin{equation}  \label{tdep}
 A^i_{CP}(f\bar{f};t)= 
 \cos(\Delta m_{\tilde{\nu}_i}t) \quad\mbox{for}\quad f\neq f'
\end{equation} 
when $\Delta \Gamma_{\tilde{\nu}} \approx 0$.
This time-dependence could be measured 
to determine the sneutrino mass splitting  
as far as $x_i$ is not too small.
On the other hand,
the time-integrated mixing probability \cite{grossman},
\begin{equation} \label{timp}
 \chi_{i} \equiv { x_{i}^2 + y_{i}^2  \over 2(x_{i}^2+1) }
\end{equation}
where $x_{i} \equiv 2\Delta m_{\tilde{\nu}_i}/
 \Gamma_{\tilde{\nu}_i}$ and $y_{i} \equiv \Delta \Gamma_{\tilde{\nu}_i}/
 \Gamma_{\tilde{\nu}_i}$.
can be used to determine the sneutrino mass splitting when $x_i \sim 1$.
Recall that $\chi_{i}$ can be measured 
by counting the `same-sign' and `opposite-sign' lepton events,
$\tilde{\nu}_i \tilde{\nu}_i^* \to  ll  \bar{l'}\bar{l'}$ and
$\tilde{\nu}_i \tilde{\nu}_i^* \to  ll' \bar{l} \bar{l'}$,
analogous to the $B$ system.
As we will see, our model generically has $y_i \ll 1$.  

For the final state $f\bar{f}$ which is shared by the decays of
$\tilde{\nu}_i$ and $\tilde{\nu}_i^*$, 
we get $X_i^{ff}=0$ and thus the time-dependent 
CP asymmetry  (for $y_i\approx 0$) becomes
\begin{equation} \label{Bcp}
 A^i_{CP}(f\bar{f};t)= {2|\rho_i|  \over 1+|\rho_i|^2 }
                  \sin(\phi_D-\phi_M) \sin \Delta m_{\tilde{\nu}_i} t 
\end{equation}
where $ \rho_i=|\rho_i|e^{i\phi_D} \equiv K_{R,i}^{f\bar{f}\,*}/
K_{L,i}^{f\bar{f}}$.
This is analogous to the CP asymmetry from $B^0$ and $\bar{B}^0$ 
decays to CP eigenstates \cite{pdg}.  

\medskip

Let us now examine how the quantities discussed above arise 
in the R-parity violating supersymmetric standard model.
For the sake of simplicity, we introduce only 
bilinear R-parity breaking terms in the superpotential $W$ 
and the soft scalar potential $V_{soft}$ of the supersymmetry 
standard model;
\begin{equation} \label{WVrp}
 W \ni -\mu_i L_i H_2, \; 
 V_{soft} \ni m^2_{iH} L_i H_1^\dagger + B_i L_i H_2 + h.c. \,
\end{equation}
where the same notations $L_i, H_{1,2}$ are used for the lepton and Higgs 
superfields and their scalar components.  This type of model is 
known to generate viable neutrino mass matrices explaining both the solar 
and the atmospheric neutrino data  \cite{hemp}.
As is  well-known, the bilinear terms in Eq.~(\ref{WVrp})
give rise to nonzero sneutrino vacuum expectation values \cite{hall};
$a_i \equiv \langle \tilde{\nu}_i^* \rangle/ \langle H_1^0 \rangle 
= (m^2_{iH} + \mu\mu_i + B_i t_\beta)/ m^2_{\tilde{\nu}_i}$
where $t_\beta=\tan\beta \equiv \langle H_2^0 \rangle/
\langle H_1^0 \rangle$ and $\mu$ is the supersymmetric Higgs mass parameter.
Then the quantities $\mu_i-\mu a_i$ determine the tree-level neutrino 
mass matrix;
\begin{equation} \label{mntree}
 m^\nu_{ij} = -M_Z^2 \xi_i \xi_j c_\beta^2/F_N
\end{equation}
where  $\xi_i \equiv a_i - \mu_i/\mu$ and 
$F_N \equiv  M_1M_2/(M_1c_W^2+M_2s_W^2)-M_Z^2\sin2\beta/\mu$.
Here $M_1$ and $M_2$ denote the $U(1)$ and $SU(2)$ gaugino masses, 
respectively.
Remember that only one neutrino, $\nu_3$, becomes  massive 
from Eq.~(\ref{mntree}), 
while the other two will get smaller masses from one-loop corrections.
The mass matrix (\ref{mntree}) fixes two neutrino mixing angles,
$\theta_{23}$ and $\theta_{13}$, corresponding to the atmospheric
neutrino and the reactor neutrino mixing angles, respectively, 
as follows \cite{vissani}:
\begin{equation} \label{s2th}
 \sin^22\theta_{23}= 4|\hat{\xi}_\mu|^2 
                     |\hat{\xi}_\tau|^2, \;
 \sin^22\theta_{13}= 4|\hat{\xi}_e|^2 (1-|\hat{\xi}_e|^2) 
\end{equation}
where $\hat{\xi}_i\equiv \xi_i/|\xi|$ and $|\xi|^2 \equiv\sum_i |\xi_i|^2$.
Current experiments require $|\xi_\mu|\approx |\xi_\tau|$ 
for the large atmospheric neutrino mixing \cite{skatm}
and $|\xi_e|/|\xi_\tau| <  0.3$
for the suppressed reactor neutrino
oscillation  $\nu_e\to \nu_{\mu,\tau}$ \cite{chooz}.

R-parity violation induces also nontrivial mixing between sneutrinos and 
neutral Higgs bosons. Furthermore, their CP-even and CP-odd parts 
mix together if general complex couplings are allowed.  
Therefore, we have to deal with a mass matrix of 10 neutral boson fields
including the Goldstone mode.  To do this, it is convenient to
define the `proper' sneutrino fields 
getting rid of the Goldstone mode by performing the see-saw rotation 
with the small angle $a_i$.  In this basis, one finds 
that the sneutrino--Higgs mixing term for each sneutrino generation
is proportional to $B_i- B a_i $ \cite{hwang}.
The next step is to rotate away these mixing terms to find 
the sneutrino mass splitting;
\begin{equation} \label{snusplit}
 \Delta m_{\tilde{\nu}_i}
 =  2 m_{\tilde{\nu}_i} M^2_Z m_A^4 |\eta_i|^2 c_\beta^2 s_\beta^4 /F_S 
\end{equation}
where  
$\eta_i \equiv a_i -B_i/B$  and 
$F_S \equiv  
     ( m^2_{\tilde{\nu}_i}-m_h^2) ( m^2_{\tilde{\nu}_i}-m_H^2)  
     (m^2_{\tilde{\nu}_i}-m_A^2) $.
Here  $m_h$ and $m_H$  denote 
the light and  heavy  CP-even Higgs boson masses, respectively and
$m_A$ is the CP-odd Higgs boson mass which are defined in the R-parity
conserving limit.  Eq.~(\ref{snusplit})  is consistent with the 
result of Ref.~\cite{haber} for the CP-conserving case.  
In deriving the above result, we neglected the sneutrino flavor mixing
induced by the neutrino flavor mixing.  This is a good
approximation as far as the mass differences between two sneutrino
flavors are much larger than $m_\nu$, $|m_{\tilde{\nu}_i}-
m_{\tilde{\nu}_j}|\gg m_\nu$, which is usually the case.  
Remark that  the universality in $B$ parameters,
$B_i/B=\mu_i/\mu$, implies $\xi_i=\eta_i$
which is not assumed in this paper.
Combining Eqs.~(\ref{mntree}) and (\ref{snusplit}), one finds
\begin{equation}\label{rsn}
 \Delta m_{\tilde{\nu}_i}/ m_{\nu_3} \sim
 (2 F_N/ m_{\tilde{\nu}_i}) (|\eta_i|^2/|\xi|^2)
\end{equation}
for $ m_{\tilde{\nu}_i} \ll m_A $.  
Therefore, we get $\Delta m_{\tilde{\nu}_i} \sim  m_{\nu_3}$ as expected
for $F_N\sim  m_{\tilde{\nu}_i}$ and $|\eta_i| \sim |\xi|$.
Here we note that the sneutrino oscillation time $1/\Delta m_{\tilde{\nu}_i}$
can be made larger than $1/m_{\nu_3}\sim 0.013$ ps for $|\eta_i| < |\xi|$. 
{}From the discussion below Eq.~(\ref{s2th}) implying 
$|\eta_e| \sim |\xi_e|\ll |\xi|$,
one finds that the electron sneutrino $\tilde{\nu}_e$ will generically
have the  largest oscillation time.
But, we will see that the $\tilde{\nu}_e$ decay to charged leptons 
have too small branching fraction to be observed.

The sneutrino couplings to light fermions $f$ and $f'$ arise from
the neutrino-neutralino, the charged lepton--chargino, 
and the slepton-Higgs mixing.  
In the bilinear model under consideration, one finds the
coupling constants $K_{L,R}$ defined in Eq.~(\ref{snuff}) as follows:
\begin{eqnarray} \label{Ks}
&&K_{R,i}^{\nu_j \bar{\nu}_k} =  
   {\delta_{ij}\over c_W} {M_Z \over F_N} \xi^*_k c_\beta,  \\
&&K_{L,i}^{\tau \bar{k}} = {m_\tau \over M_W} 
[\delta_{i\tau} {\omega_{k1}\over c_\beta} + 
            \delta_{k\tau} {\theta_{i1}\over c_\beta} ] \nonumber
\,,  \quad 
K_{L, i}^{b\bar{b}} = {m_b \over M_W} {\theta_{i1}\over c_\beta}\,,  \\
&&K_{R,i}^{j \bar{\tau}} = {m_\tau \over M_W} 
[\delta_{ij} {\omega_{\tau2}^* \over c_\beta}
 + \delta_{j\tau} {\theta_{i2}^*\over c_\beta} ],
\quad
K_{R,i}^{b\bar{b}} = {m_b \over M_W} {\theta_{i2}^*\over c_\beta} , \nonumber
\end{eqnarray}
where the coefficients $\omega$ and $\theta$
are given by
\begin{eqnarray} \label{thetas}
&&\omega_{k1} = 
 { -\mu M_2\, \xi_k\over \mu M_2-M_W^2s_{2\beta} } + a_k,\; 
\omega_{k2} =  
 {2 \mu  (\mu + M_2 t_\beta) M_W^2 \xi_k\over (\mu M_2-M_W^2s_{2\beta})^2 }
  c_\beta^2,  \nonumber\\
&&\theta_{i1} = - a_i - s^2_\beta[ m^4_{\tilde{\nu}_i} - 
m^2_{\tilde{\nu}_i} m^2_A-(m^2_{\tilde{\nu}_i}+ m_A^2c_{2\beta})
M_Z^2s^2_\beta] m_A^2 \eta_i/F_S
 \nonumber \\
&&\theta_{i2} = - c^2_\beta s^2_\beta M_Z^2 [ m^2_{\tilde{\nu}_i}
 - m_A^2 c_{2\beta} ] m_A^2 \eta_i/F_S  
 \,.  
\end{eqnarray}
In Eq.~(\ref{Ks}),  we neglected the terms proportional to 
lighter quark and lepton masses.  
In the presence of trilinear R-parity violating terms,
the couplings (\ref{Ks}) will get additional contributions leaving
Eqs.~(\ref{mntree}) and (\ref{snusplit}) unchanged.

\medskip

We are ready to discuss how the observable sneutrino mixing phenomena
occur in our scheme.
{}From Eqs.~(\ref{Ks}) and (\ref{thetas}), one can see that
the dominant decay channels are $\tilde{\nu}_i \to \nu\bar{\nu}$ and 
$\tilde{\nu}_i \to b\bar{b}, \tau\bar{\tau}$ for low and 
large $\tan\beta$, respectively, if the R-parity violating parameters
$\omega_i, \theta_i$, are of the same order.
For the latter decay channels, one also finds that $|K_L|$ is larger 
than $|K_R|$ for the parameters $M_Z<M_2,\mu$ and large $\tan\beta$.
This shows that we generally have $y=\Delta\Gamma/\Gamma \ll 1$ 
as the decay rate difference $\Delta \Gamma$ gets non-vanishing contributions 
from $K_L^{f\bar{f}} K_R^{f\bar{f}}$ where $f=\tau$, $b$.
Since we are interested in the region where
$ \Gamma_{\tilde{\nu}}<  \Delta  m_{\tilde{\nu}}$, 
we can put $c(t)=1$ and $s(t)=0$ in Eq.~(\ref{Acpt}).
Let us now make an estimate of the quantity $x_i$.
Taking $\Gamma_{\tilde{\nu}_i}$ dominated by the couplings 
$K_{R,i}^{\nu\bar{\nu}}$ or $K_{L,i}^{\tau\bar{\tau}}$, we get 
\begin{equation} \label{xis}
x_{i} \sim
{8 \over \alpha_2} \cdot {\rm Min}\left[
{F_N^2\over m_{\tilde{\nu}_i}^2} 
\left|{\eta_i\over \xi}\right|^2,\;
{1\over t_\beta^2} {(M_Z M_W)^2 \over 
(m_{\tilde{\nu}_i} m_\tau t_\beta)^2}
\left|\eta_i \over \theta_{i1}\right|^2 
\right]
\end{equation}
which shows $x_i \gg 1$ for small $\tan\beta$.
With $m_{\tilde{\nu}_i}=M_Z=F_N/2$ and $|\eta_i|=|\xi|=|\theta_{i1}|$, 
one gets $x_i \sim 1$ with $\tan\beta\sim 20$ for which the time-integrated
quantities can be well measured.  On the other hand, for low $\tan\beta$,
the time-dependent observables may be measured to determine
$\Delta m_{\tilde{\nu}}$ or CP asymmetry.
If  $\tan\beta$ is too large ($x_i \ll 1$), no oscillation effect occurs.
The quantity $\rho_i$ determining the magnitude of CP asymmetry is given by
\begin{equation} \label{rhos}
\rho_i = \theta_{i2}/\theta_{i1} 
\;\;{\rm or}\;\;
(\delta_{i\tau}\omega_{i2}+\theta_{i2})/ 
(\delta_{i\tau}\omega_{i1}+\theta_{i1}) 
\end{equation}
for the shared final states $b\bar{b}$ or $\tau \bar{\tau}$, respectively.
As can be inferred from previous discussions, 
the magnitude of $\rho_i$ is usually smaller than 1 
and drops down with growing $\tan\beta$.
This should be compared with the quantity $x_i$ which shows the
similar behavior as above.  Thus, we typically get $|\rho_i| \ll 1$ 
in the region where $x_i\sim 1$.  
For the measurement of the CP asymmetries in the shared final states,
one should determine whether the decaying particle is a sneutrino or an
anti-sneutrino.  Obviously, this can be done by looking at the
flavor-specific decays of the other particle, for instance, 
$\tilde{\nu}_{\alpha} \to \alpha\bar{\tau} 
\leftarrow\hspace{-2.5ex}/\hspace{1ex} \tilde{\nu}_{\alpha}^*$  and 
$\tilde{\nu}_{\tau} \to \tau\bar{\alpha} 
\leftarrow\hspace{-2.5ex}/\hspace{1ex} \tilde{\nu}_{\tau}^*$
where $\alpha=e,\mu$. Therefore,
the branching fractions for the flavor-specific 
as well as the shared final states should be large enough.  
To get a reference value, let us consider a future $e^+ e^-$ linear collider 
with an integrated luminosity 1000 fb$^{-1}$ at $\sqrt{s}=500$ GeV.  
For the cross-section of the sneutrino--anti-sneutrino pair production 
$\sim$ 50 fb with $m_{\tilde{\nu}} \sim 100$ GeV \cite{bartle}, 
there will be about $5\times10^4$ events.  
Requiring the multiplied branching ratio to be larger than 1 \%,
one will get about 500 samples to study CP violation.
Here, we note that such a value can be hardly obtained for
the electron and muon sneutrinos.  This is because
the decay $\tilde{\nu}_\alpha \to \alpha \bar{\tau}$ ($\alpha=e,\mu$) comes
from the coupling $K^{\alpha\bar{\tau}}_{R,\alpha}$ which is suppressed
by the factor $m_\tau/\tan\beta$.
Considering the ratio of $K^{\alpha\bar{\tau}}_{R,\alpha}$ to
$K^{\nu\bar{\nu}}_{R,\alpha}$ and $K^{\tau\bar{\tau}}_{L,\alpha}$,
we get the branching fraction,  BR($\alpha\bar{\tau}$) 
$=| K^{\alpha\bar{\tau}}_{R,\alpha}/K^{\nu\bar{\nu}}_{R,\alpha}|^2
\approx (m_\tau t_\beta M_W/ \mu M_Z)^2 $ for low $\tan\beta$, and 
BR($\alpha\bar{\tau}$) 
$= |K^{\alpha\bar{\tau}}_{R,\alpha}/K^{\tau\bar{\tau}}_{L,\alpha}|^2
\approx (M_W^2/ t_\beta \mu M_2)^2$ for large $\tan\beta$,
which  shows that the branching ratio is usually smaller than
1 \% for $M_W < M_2\sim F_N \sim \mu$.

However, the situation can be different for the tau sneutrino
which allows 
${\rm BR}(\tau\bar{\mu}) \sim 
 {\rm BR}( \tau\bar{\tau}) \sim 
 {\rm BR}( b\bar{b}) \gg 
 {\rm BR}(\nu\bar{\nu})  $
for $|\omega_{\mu 1}|\sim |\omega_{\tau 1}| \sim |\theta_{\tau 1}|$ and 
relatively large $\tan\beta$, as the relevant coupling 
$K^{\tau\alpha}_{L,\tau}$ has no such suppression.  
To show explicitly 
the sneutrino oscillation and corresponding CP violation effects,
we choose the following typical set of mass parameters as a reference: 
\begin{equation} \label{sett}
  M_2=\mu=2 m_L,\; m_A=2.5 m_L,\; m_h=115\,{\rm GeV}
\end{equation}
where $m_L$ denotes the slepton doublet soft mass.  
Taking  $m_L=100$ (200) GeV, the sneutrino masses are,
$m_{\tilde{\nu}}= \sqrt{m_L^2+M_Z^2 c_{2\beta}/2}
=$ 82, 77 and 76 (192, 189 and 189) GeV 
for $\tan\beta=3, 15$ and 30, respectively.
Concerning the gaugino masses, the unification relation is assumed; 
$M_2 = 2 M_1$.   
In our bilinear model, there are three types of R-parity violating 
parameters $\mu_i$, $B_i$ and  $m^2_{iH}$ 
without assuming the universality.
For our calculation, we will trade these parameters with 
$\xi_i$, $\eta_i$ and $a_i$. Then, we  put
$\xi_\mu=\xi_\tau=3\xi_e$, consistently with the experimental data 
as mentioned before and the value of $\xi_\tau$ is normalized to yield 
$m_{\nu_3}=0.05$ eV in accordance with Eq.~(\ref{mntree}).  
In fact, our results are insensitive to the ratio 
$\xi_e/\xi_\tau$.  
Note that small neutrino masses requires very small R-parity violating 
numbers, $\xi_i c_\beta \sim 10^{-6}$,
which could be a consequence of some flavor symmetry explaining 
quark and lepton Yukawa hierarchies \cite{u1}.  

To find the favorable parameter space, we consider the following 
two regions: (a) $|\xi_\tau| \sim |\eta_\tau| \sim |a_\tau|$ and 
(b) $|\xi_\tau|  \gg |\eta_\tau| \sim |a_\tau|$.
As discussed  before, the latter region is chosen to give 
larger $|\xi_\tau|$ and thus larger oscillation time.
In the first region, the oscillation time becomes too small 
for the time-dependence to be observed.
In Table A and B, we present
the branching ratios and various oscillation parameters for
the tau sneutrino LSP taking the following specific numbers 
for the two regions:
\begin{equation} \label{AB}
\mbox{(A)}\;\; \xi_\tau= -\eta_\tau = -2 a_\tau, \qquad  
\mbox{(B)}\;\; \xi_\tau=6 \eta_\tau = -6a_\tau.
\end{equation}
{}From Table A, one finds that $x$ becomes of order one 
for $\tan\beta \sim 30$  and the proper oscillation time $t_{osc}$
is very small.  As the branching ratios for 
$\tilde{\nu}_\tau \to \tau\bar{\tau}, \tau\bar{\mu}$ become 40--50\%
for large $\tan\beta$, one could look for the same-sign lepton signal to
measure the time-integrated mixing probability $\chi$. 
In this case, the CP asymmetries $\rho$ are smaller than 1 \% and thus
can be hardly measured.
As alluded before, the set (B) gives  observable time-dependence.
As can be seen from Table B, 
$\rho$ becomes smaller and the branching ratios larger 
as $\tan\beta$ grows.  The multiplied branching ratios for the leptonic modes
becomes of order 10\% for intermediate to large $\tan\beta$ and thus
the sneutrino mass splitting can be directly obtained
by tracing time-dependent oscillation for the final state
$\tau\bar{\mu}$ in this region.
Furthermore,  the time-dependent CP asymmetry for the final state 
$\tau\bar{\tau}$ can be a few  percent  assuming 
the  maximal CP phase $|\sin(\phi_D-\phi_M)| \sim 1$, which
could be within the future experimental reach.
In fact, the quantity $\rho$ can be made larger for $|\xi|\gg |\eta_i|$
and lower $\tan\beta$. However, restricting ourselves to
the region of $|\xi| < 10 |\eta_i|$, we 
find that $\rho$ can be maximally a few \%.  
Therefore, a better luminosity than mentioned before will be needed 
to cover more parameter space of our model.
We note here that one hardly gets large CP asymmetry for 
$\tilde{\nu}_\tau \to b\bar{b}$ as the corresponding $\rho$ is
proportional to $t^{-2}_\beta$.  
In both cases of (A) and (B), one finds that smaller $m_L$ is
favored for the observation of various sneutrino oscillation observables.

Let us finally comment on the case of the universal soft parameters
giving $\xi_i \approx \eta_i \ll a_i$.  
{}From Eq.~(\ref{Ks}) and (\ref{thetas}), one can see that 
the sneutrino decays dominantly to $b\bar{b}$ and the branching fraction
for $\tau\bar{\mu}$ is $m_\tau^2/3m_b^2 \sim$ 5\% with
much more suppressed rate for $\tau\bar{\tau}$.  Thus, almost no 
oscillation effects can be observed in this case.

\medskip

In conclusion, R-parity violation can lead not only to the realistic
neutrino masses and mixing,  but also to observable sneutrino--anti-sneutrino 
mixing phenomena.  This can occur if the tau sneutrino is the LSP.
Among various observable quantities in the sneutrino oscillation, 
a certain observable will only be within 
experimental reach given parameter region of the bilinear model of
R-parity violation.
For the less fine-tuned bilinear R-parity violating parameters,
$|\xi|\sim|\eta|$, only time-integrated quantities can be probed, whereas
the time-dependence and CP asymmetries in the sneutrino oscillation 
can be observed for the region where $|\xi|\gg |\eta|$. 
In the former region,
the sneutrino mass splitting  can be determined by measuring  
the time-integrated mixing probability for large 
$\tan\beta$ where $x$ becomes order one.  
In the latter region,  the mass splitting can be determined through 
time-dependent oscillation even for lower  $\tan\beta$.
Finally, if the sneutrino mass is small ($m_{\tilde{\nu}}< 100$ GeV)
and $\tan\beta$ is in the intermediate region,
the CP asymmetry in the decay $\tilde{\nu}_\tau \to \tau\bar{\tau}$ 
can reach the level of a few percent
which could be measured in the future experiments.

\medskip

{\bf Acknowledgement}:
The author thanks JoAnne Hewett for discussions and   
SLAC Theory Group for its hospitality.
This work is supported by
BK21 program of the Ministry of Education.


\newpage
\begin{table} 

\begin{tabular}{c|c||c|c|c|c||c|c|c|c|c} \hline
$m_L$/GeV & $\tan\beta$ & $\nu\bar{\nu}$ & $\tau\bar{\tau}$
       & $\tau\bar{\mu}$ & $b\bar{b}$ & 
  $x$ & $y$ & $\rho(\tau\bar{\tau})$ & $\rho(b\bar{b})$ & $t_{osc}$/ps 
\cr \hline
&3       & $0.68$  & $0.03$ & $0.07$ & $0.22$ & 
         13035 & 0.08  & 0.037 & 0.20 & 0.008   
\cr
100
&15 & $0.007$ & $0.38$  & $0.47$  & $0.09$ & 
         197   & 0.02 & 0.013 & 0.04 & 0.007  
 \cr
&30      & 5$\times$$10^{-4}$ & $0.40$ & $0.48$ & $0.08$ & 
         13  & 0.007 & 0.007 & 0.012& 0.007  
\cr
\hline
&3       & $0.17$ & $0.21$ & $0.084$ & $0.53$ & 
         12670 & 0.062 & 0.031  & 0.047 & 0.0048  
\cr
200
&15 & 4$\times$$10^{-4}$ & $0.24$ & $0.092$ & $0.66$ & 
         35   & 0.005 & 0.003 & 0.003 & 0.0037  
\cr
&30      & 2$\times$$10^{-5}$ & $0.24$ & $0.092$ & $0.66$ & 
         2.2   & 0.002 & 0.001 & 0.0007 & 0.0037  
\cr
\hline
\end{tabular}

\vspace{2ex}
{Table A:  
The branching ratios of the tau sneutrino decay into $\nu\bar{\nu}$,
$\tau\bar{\tau}$, $\tau\bar{\mu}$ and $b\bar{b}$ are shown together with
the sizes of the oscillation parameters $x,y$ and $\rho$ 
defined in Eqs.~(\ref{timp}) and (\ref{Bcp}), respectively.  
The choice of R-parity conserving parameters is described in Eq.~(\ref{sett}) 
and the set (A) in Eq.~(\ref{AB}) is used
for the R-parity violating parameters.  }

\vspace{1cm}
\begin{tabular}{c|c||c|c|c|c||c|c|c|c|c} \hline
$m_L$/GeV & $\tan\beta$ & $\nu\bar{\nu}$ & $\tau\bar{\tau}$
       & $\tau\bar{\mu}$ & $b\bar{b}$ & 
  $x$ & $y$ & $\rho(\tau\bar{\tau})$ & $\rho(b\bar{b})$ & $t_{osc}$/ps 
\cr \hline
&3       & $0.87$  & $0.05$ & $0.06$ & $0.01$ & 
         468 & 0.01 &  0.16  & 0.14  & 0.30   
\cr
100
&15 & $0.011$ & $0.27$  & $0.41$  & $0.28$ & 
         7.7   & 0.011 & 0.026 & 0.005 & 0.25  
\cr
&30      & 7$\times$$10^{-4}$ & $0.27$ & $0.41$ & $0.29$ & 
         0.50  & 0.006 & 0.012  & 0.001 & 0.24  
\cr
\hline
&3       & $0.48$ & $0.06$ & $0.14$ & $0.30$ & 
         985 & 0.015 & 0.034  & 0.017 &  0.17 
\cr
200
&15 & 2$\times$$10^{-3}$ & $0.11$ & $0.26$ & $0.60$ & 
         4.6   & 0.003 & 0.007  & 0.001 & 0.13  
\cr
&30      & 1$\times$$10^{-4}$ & $0.11$ & $0.26$ & $0.60$ & 
         0.29   & 0.001 & 0.004  & 0.0003 & 0.13  
\cr
\hline
\end{tabular}

\vspace{2ex}
{Table B:  The same as above but with the set (B) in Eq.~(\ref{AB}). }

\end{table}

\end{document}